\documentclass[apjl,twocolumn,floats]{emulateapj}
\pdfoutput=1
\usepackage{bm}
\usepackage{multirow}
\usepackage{color}
\usepackage{hyperref}
\bibpunct[; ]{(}{)}{;}{a}{}{;}

\usepackage{amsmath}
\usepackage{amsfonts}
\usepackage{amssymb}
\usepackage{epsfig}
\usepackage{hyperref}
\usepackage{graphicx}

\newcommand{\n}{{\bf{n}}}

\newcommand{\wjjj}[6]
{{
\left(
\begin{array}{lcr} #1 & #2 & #3 \\#4 & #5 & #6 \end{array}
\right)
}}

\newcommand{\wjjjjjj}[6]
{{
\left\{
\begin{array}{lcr} #1 & #2 & #3 \\#4 & #5 & #6 \end{array}
\right\}
}}

\newcommand{\bn}{\hat{\bf n}}

\newcommand{\perm}{\ {\rm perm.}}

\newcommand{\beq}{\begin{equation}}
\newcommand{\eeq}{\end{equation}}
\newcommand{\beqa}{\begin{eqnarray}}
\newcommand{\eeqa}{\end{eqnarray}}
            
\usepackage{epstopdf}
\newcommand{\wjm}{\left(
                         \begin{array}{ccc}
       l & l'  & L  \\
         m & -m'  & -M
                         \end{array}
                   \right)}

\begin{document}

\title{A Constraint On the Integrated Mass Power Spectrum out to \MakeLowercase{\textit{z}}  = 1100 \\ from Lensing of the Cosmic Microwave Background}
\author{ Joseph Smidt$^{1}$, Asantha Cooray$^{1}$, Alexandre Amblard$^{1}$, Shahab Joudaki$^{1}$, Dipak Munshi$^{2,3}$, Mario G. Santos$^{4}$, Paolo Serra$^{1}$}
\affiliation{$^{1}$Center for Cosmology,
Department of Physics  and  Astronomy,
University  of California, Irvine, CA 92697}
\affiliation{$^{2}$Scottish Universities Physics Alliance,~ Institute for Astronomy,
University of Edinburgh, Blackford Hill,  Edinburgh EH9 3HJ, UK}
\affiliation{$^{3}$School of Physics and Astronomy, Cardiff University, CF24 3AA}
\affiliation{$^{4}$CENTRA, Departamento de F\'{i}sica, Instituto Superior Tecnico, 1049-001 Lisboa, Portugal.}

\date{\today}

\begin{abstract} 
The temperature fluctuations and polarization of the Cosmic Microwave Background (CMB) are now a well-known probe of the Universe at an infant age of 400,000 years. During the transit to us from the surface of last scattering, the CMB photons are expected to undergo modifications induced by the intervening large-scale structure. Among the expected secondary effects is the weak gravitational lensing of the CMB by the foreground dark matter distribution.  We derive a quadratic estimator that uses the non-Gaussianities generated by the lensing effect at the four-point function level to extract the power spectrum of lensing potential fluctuations integrated out to $z \sim 1100$ with peak contributions from potential fluctuations at $z$ of 2 to 3. Using WMAP 7-year temperature maps, we report the first direct constraints of this lensing potential power spectrum and find that it
has an amplitude of $A_L = 0.96 \pm 0.60$, $1.06 \pm 0.69$ and $0.97 \pm 0.47$
using the W, V and W+V bands, respectively.
\end{abstract}
\keywords {cosmology: cosmic microwave background --- cosmology: observations --- cosmology: theory --- gravitational lensing}

\maketitle

\section{Introduction}

Measurements of Cosmic Microwave Background (CMB) anisotropies have served as the strongest experimental probe of the early Universe to date (Spergel et al. 2003). 
Temperature fluctuations capture the physics of the primordial photon-baryon fluid undergoing oscillations in potential wells sourced by primordial density perturbations~(Hu et al. 1997; Hu \& Dodelson 2002). The anisotropies in the CMB also contain information related to the Universe at late times, as CMB photons in transit to us encounter large-scale structure. The main mechanisms that affect the frequency and direction of propagation of CMB photons are the gravitational interaction with time varying potential perturbations~(Sachs \& Wolfe 1967; Rees \& Sciama 1968), and Compton scattering by electrons due to the reionization of the universe~(Sunyaev \& Zel'dovich 1970). 
These mechanisms create specific signatures in the CMB that can be used to extract properties of the large-scale structure. 

One observational signature present in the CMB is the gravitational lensing modification by the projected dark matter distribution integrated along the line of sight  (Hu 2000; Lewis \& Challinor 2006). Thi is analogous to weak lensing of galaxy shapes, now understood to be a strong probe of the dark matter distribution~(Wittman et al. 2000; Bacon et al. 2000, van Waerbeke et al. 2000;  Bartelmann \& Schneider 2001).  
Unlike the case of lensing measurements with galaxy shapes, which restrict studies out to the era of $z \sim 2$,  lensing of the CMB traces all the way back to the surface of last scattering at $z \sim 1100$.   The dominant contributions to CMB lensing arise from $z \sim2$, but with a 30\% contribution at higher redshifts  (Lewis \& Challinor 2006).
From the ability to probe distance ratios and the integrated matter power spectrum at early times, CMB lensing is understood to be a powerful probe of certain cosmological parameters such as the
neutrino mass and early dark energy~(Kaplinghat et al. 2003, Smith et al. 2008, Lesgourgues et al. 2006, Li \& Cooray 2006).  

Unlike lensing measurements with discrete galaxies, the primordial CMB sky is a continuous field, and a different technique is therefore needed. Due to lensing deflections, the temperature anisotropy $\Theta({\bf \hat n}) \equiv \Delta T/T({\bf \hat n})$ measured on the sky 
becomes $\tilde \Theta({\bf \hat n}+{\bf \alpha})$, where ${\bf \alpha} = \nabla \phi({\bf \hat n})$ is the deflection angle due to lensing 
given by the angular gradient of the lensing potential. The lensing potential is the line of sight projection between us and the last scattering surface of 
the gravitational potential $\Phi(r,\bn r)$ in the Universe:
\begin{equation}
\phi(\bn) = -2 \int_0^{r_0}dr {d_A(r_0-r) \over d_A(r) d_A(r_0)}\Phi(r,\bn r),
\end{equation}
where $d_A$ is the comoving angular diameter distance and $r$ is the comoving conformal distance from the observer.  Using this expression, we can consider the dominant correction to the CMB temperature as a term that couples the deflection angle $\nabla \phi(\bn)$ to the gradient of the temperature:
\begin{equation}
\label{eq:pert}
\Theta({\bf \hat n}) = \tilde \Theta({\bf \hat n + \alpha}) \sim \tilde \Theta({\bf \hat n}) + \nabla_i \phi({\bf \hat n}) \nabla^i \tilde \Theta({\bf \hat n}) ,
\end{equation}
where $\tilde \Theta({\bf \hat n})$ is the unlensed temperature map. 

The CMB lensing effect is a modification to the angular gradient of temperature on the sky. If the CMB were completely isotropic, lensing modifications would not leave a change as lensing conserves surface brightness. 
Further, the lensing signatures are at second-order in temperature leading to
a distinct non-Gaussianity pattern at the four-point function level, or trispectrum in Fourier space~(Zaldarriaga 2000; Hu 2001b). 
Quadratic statistics can be devised to probe the gradient structure of the CMB temperature 
and to extract the projected line of sight dark matter density field~(Hu 2001a; Cooray \& Kesden 2003; Okamoto \& Hu 2003).  
Attempts have been made for an indirect signature of the lensing effect in all-sky maps by Wilkinson Microwave Anisotropy Probe (WMAP) three-year
data by estimating the deflection field and then cross-correlating that with a foreground density tracer field, such as NRAO VLA Sky Survey (NVSS) and Sloan Digital Sky Survey (SDSS). The overall signal-to-noise ratio for the measurement is 2.5$\sigma$~(Hirata et al. 2008) to 3.4$\sigma$~(Smith et al. 2007).

With observations over seven years now completed, WMAP maps have improved in sensitivity to the extent that a direct measurement of the lensing signal can be pursued.
We first construct an optimized statistic that probes the non-Gaussianity pattern induced by lensing via the trispectrum of the CMB temperature. We measure this by four maps weighted differently and taking a power spectrum of the temperature squared field.
In addition to the lensing signal, this power spectrum contains a term associated with the Gaussian sky. We combine the constraints in data with a large suite of simulations, involving both the Gaussian temperature maps without lensing as well as maps with lensing included.

This {\it Letter} is organized as follows. In the next Section, we outline the new estimator and how to weight CMB maps to extract the lensing signal.  In Section~\ref{sec:test} we show that our estimator works correctly for simulated data.  In Section~\ref{sec:Analysis} we describe our analysis used to constrain  $C_l^{\phi \phi}$, the detection of the lensing amplitude and describe null tests used as a sanity checks.  In Section~\ref{sec:final} we present our results and conclude with a discussion on implications for the future.

\section{Derivation of Estimator}
\label{sec:derive}

We derive our estimator using the trispectrum of the CMB. (Hu 2001b; Okamoto \& Hu 2003; Smidt et al. 2010).   In harmonic space, the lensing perturbed temperature field becomes (Eq.~\ref{eq:pert}):
\begin{eqnarray}
\label{eq:dtheta}
\delta \Theta_{lm} &\sim&  \int d\n Y_l^{m*}  \nabla_i \phi({\bf \hat n}) \nabla^i \tilde \Theta({\bf \hat n}) \\
&=& \int d\n Y_l^{m*}  \nabla_i \left( \sum_{LM}  \phi_{LM} Y_L^{M} \right) \nabla^i  \left( \sum_{l' m'} \tilde \Theta_{l'm'} Y_{l'}^{m'} \right) \nonumber \\
&=& \sum_{LM} \sum_{l' m'} \phi_{LM} \tilde \Theta_{l'm'}(-1)^m \wjm F_{l l' L}, \nonumber
\end{eqnarray}
with
\beqa
F_{l l' L} = \sqrt{(2l+1) (2l'+1) (2L+1) \over 4 \pi}  \wjjj{l}{l'}{L}{0}{0}{0} \\
\times {1 \over 2} [L(L+1) + l'(l'+1) - l(l+1)], \nonumber
\eeqa
where the term in large parenthesis is the Wigner-3j symbol.

From here, it is straightforward to show that lensing does not generate a three
point correlation function or a bispectrum. The first correction to non-Gaussianity from CMB lensing appears in the four point correlation function or, in Fourier space, the angular trispectrum:
\begin{eqnarray}
&& \left< \Theta_{l_1 m_1} \Theta_{l_2 m_2} \Theta_{l_3 m_3} \Theta_{l_4 m_4}\right> =   \\
 && \left< \Theta_{l_1 m_1} \Theta_{l_2 m_2} \Theta_{l_3 m_3} \Theta_{l_4 m_4}\right>_{\rm G}+  \left< \Theta_{l_1 m_1} \Theta_{l_2 m_2} \Theta_{l_3 m_3} \Theta_{l_4 m_4}\right>_{\rm c}. \nonumber 
 \end{eqnarray}
 where the first term on the right hand side represents the contribution from a Gaussian sky and is subtracted out in our analysis (Hu 2001b, Smidt et al. 2010).  The second term is called the connected piece containing the contribution to the trispectrum from lensing and is expanded as 
 \begin{eqnarray}
&& \left< \Theta_{l_1 m_1} \Theta_{l_2 m_2} \Theta_{l_3 m_3} \Theta_{l_4 m_4}\right>_{\rm c} = \\
&& \hspace{0.3 cm} \sum_{L M} (-1)^M P^{l_1 l_2}_{l_3 l_4}(L) 
\wjjj{l_1}{l_2}{L}{m_1}{m_2}{-M} \wjjj{l_3}{l_4}{L}{m_3}{m_4}{M}  \nonumber \\
&& \hspace{0.3 cm} + (l_2 \leftrightarrow l_3) + (l_2 \leftrightarrow l_4) \nonumber 
\end{eqnarray}
where 
\begin{eqnarray}
\label{eq:pair}
P^{l_1 l_2}_{l_3 l_4}(L) &&= \\
&& \hspace{-1.5cm}C_L^{\phi \phi} \left(  \tilde C_{l_2} F_{l_1 l_2 L}+  \tilde C_{l_1} F_{l_2 l_1 L} \right)  \left(  \tilde C_{l_4} F_{l_3 l_4 L}+  \tilde C_{l_3} F_{l_4 l_3 L} \right) \nonumber
\end{eqnarray}
Here $P^{l_1 l_2}_{l_3 l_4}(L)$ is known as a ``pairing" of the full trispectrum (Hu 2001b).

From this, we can write the full trispectrum $T^{l_1 l_2}_{l_3 l_4}(L)$ keeping all the permutations as:
\begin{eqnarray}
&& T^{l_1 l_2}_{l_3 l_4}(L) = P^{l_1 l_2}_{l_3 l_4}(L) + \left( 2L+1\right)\sum_{L'}  \bigg[(-1)^{l_2+l_3} \wjjjjjj{l_1}{l_2}{L}{l_4}{l_3}{L'} \nonumber \\
&& \times P^{l_1 l_3}_{l_2 l_4}(L') + (-1)^{L+L'} \wjjjjjj{l_1}{l_2}{L}{l_3}{l_4}{L'} P^{l_1 l_4}_{l_3 l_2}(L') \bigg] 
\end{eqnarray}
where the curly bracket is the Wigner 6-j symbol.

It has been shown that for weak lensing of the CMB, the full trispectrum is approximated well by its pairing 
and thus we will approximate the full trispectrum as  $T^{l_1 l_2}_{l_3 l_4}(L) = P^{l_1 l_2}_{l_3 l_4}(L)$~(Hu 2001b).
This simplifies our analysis significantly and we have verified that this approximation is accurate to better than 3\%, which is adequate given our overall constraints are only accurate to 25\%.

To extract $C_l^{\phi \phi}$ from CMB maps, we develop a quadratic estimator based on the kurtosis power spectrum of the CMB. Starting with Eq.~\ref{eq:pair}, we find the trispectrum is of the form:
\beqa
\label{eq:Tperm}
T_{l_1l_2}^{l_3l_4}(L) &=& {1 \over 4}h_{l_1 l_2 L} h_{l_3 l_4 L} C_l^{\phi \phi} \nonumber \\
&\times&  \left( \tilde C_{l_2}  \tilde C_{l_4} I_{l_1 l_2 L} I_{l_3 l_4 L}+  (3 \perm) \right) 
\eeqa
where
\beqa
h_{l_1 l_2 L} &=& \sqrt{(2l_1+1) (2l_2+1) (2L+1) \over 4 \pi}  \wjjj{l_1}{l_2}{L}{0}{0}{0} \nonumber \\
I_{l_1 l_2 L} &= & [L(L+1) + l_2(l_2+1) - l_1(l_1+1)].
\eeqa

This lets us break up $T_{l_1l_2}^{l_3l_4}(l)$ into $4\times9 = 36$ distinct pieces:
\beq
\label{eq:twelve}
T_{l_1l_2}^{l_3l_4}(l) = T_{l_1l_2}^{(1) l_3l_4}(l) +T_{l_1l_2}^{(2) l_3l_4}(l) +...+T_{l_1l_2}^{(36) l_3l_4}(l) ,
\eeq
allowing us to write ${\cal K}_{l}^{2,2}$ as
\beqa
\label{eq:expK}
 && {\cal K}_{l}^{(2,2)} =  {1 \over (2l+1)} \\
  && \times \sum_{l_i} {1 \over (2l+1)}
 { \left(\hat T_{l_1l_2}^{(1)l_3l_4}(l) +...+\hat T_{l_1l_2}^{(36)l_3l_4}(l) \right)T^{l_1l_2}_{l_3l_4}(l) \over {\cal C}_{l_1} {\cal C}_{l_2} {\cal C}_{l_3} {\cal C}_{l_4}} \nonumber. 
\eeqa
Here ${\cal C}_{l} = C_l b_l^2 + N_l$ where $C_l$ is the temperature power spectrum, $b_l$ is the beam transfer function related to the experiment,  $N_l$ is the noise power spectrum is the temperature power spectrum and $T^{l_1l_2}_{l_3l_4}(l)$ is the full trispectrum.   Also, $ \hat T^{l_1l_2}_{l_3l_4}(l)$ is the full trispectrum with $C_l^{\phi \phi} = 1$ leaving the righthand side proportional to $C_l^{\phi \phi}$.  For convenience we introduce the notation ${\cal K}_{l}^{(2,2)} = C_l^{\phi \phi} K_{l}^{(2,2)}$.

Each term of Eq.~\ref{eq:twelve}, $\hat T_{l_1l_2}^{(x) l_3l_4}(l)$, can be broken up into the form
\beq
\label{eq:abgd}
\hat T_{l_1l_2}^{(x) l_3l_4}(L) = h_{l_1 l_2 L} h_{l_3 l_4 L} F_L \alpha^{(x)}_{l_1} \beta^{(x)}_{l_2} \gamma^{(x)}_{l_3} \delta^{(x)}_{l_4}.
\eeq
This allows us to recover $ {\cal K}_{l}^{(2,2)}$ from weighted CMB maps analogously to what was done for primordial non-Gaussianity by Munshi et al. (2009) and Smidt et al. (2010).

First we introduce the necessary weighted CMB maps:
\begin{eqnarray}
\label{eq:Amap}
A^{(x)}(\n) &\equiv& \sum_{lm} Y_{lm}(\n) A_{lm}^{(x)}; ~ A_{lm}^{(x)} \equiv {\alpha_{l}^{(x)} \over  {\cal C}_l} b_l a_{lm} \\
B^{(x)}(\n) &\equiv& \sum_{lm} Y_{lm}(\n) B_{lm}^{(x)}; ~ B_{lm}^{(x)} \equiv {\beta_{l}^{(x)} \over {\cal C}_l} b_l a_{lm} \\
G^{(x)}(\n) &\equiv& \sum_{lm} Y_{lm}(\n) G_{lm}^{(x)}; ~ G_{lm}^{(x)} \equiv {\gamma_{l}^{(x)} \over  {\cal C}_l} b_l a_{lm} \\
\label{eq:dmap}
D^{(x)}(\n) &\equiv& \sum_{lm} Y_{lm}(\n) D_{lm}^{(x)}; ~ D_{lm}^{(x)} \equiv {\delta_{l}^{(x)} \over {\cal C}_l} b_l a_{lm}. 
\end{eqnarray}
Here, the specific $\alpha_l^{(x)}, \beta_l^{(x)}, \gamma_l^{(x)}$ and $\delta_l^{(x)}$ weightings for each of the 36 terms of the trispectrum may be deduced from Eq.~\ref{eq:abgd} and $a_{lm}$ comes from the CMB map being weighted.

From these maps we can form a quadratic statistic for each term
\beqa
\label{eq:sumk22}
 {\cal K}_{l}^{(x)(2,2)} &=& {\cal K}_{l}^{A^{(x)} B^{(x)}, G^{(x)} D^{(x)}},\\
&=& {1 \over (2l+1)} \sum_m \left[A^{(x)}B^{(x)}\right]_{lm} \left[G^{(x)}D^{(x)}\right]_{lm}, \nonumber
\eeqa
where $[A^{(x)}B^{(x)}]_{lm}$ are the spherical harmonics from the product of the 
$A^{(x)}$ and $B^{(x)}$ maps with $[G^{(x)}D^{(x)}]_{lm}$ defined similarly.

From these definitions it is straightforward to recover:
\beq
{\cal K}_{l}^{A^{(x)} B^{(x)}, G^{(x)} D^{(x)}}  =  {1 \over (2l+1)} \sum_{l_i} {1 \over (2l+1)}
 { \hat T_{l_3l_4}^{(x)l_1l_2}(l) T^{l_1l_2}_{l_3l_4}(l) \over {\cal C}_{l_1} {\cal C}_{l_2} {\cal C}_{l_3
} {\cal C}_{l_4}},
\eeq
so that using Eq.~\ref{eq:twelve} we indeed recover  Eq.~\ref{eq:expK} noting
\beq
\label{eq:eexpK}
{\cal K}_{l}^{(2,2)} = {\cal K}_{l}^{(1)(2,2)}+{\cal K}_{l}^{(2)(2,2)}+...+{\cal K}_{l}^{(36)(2,2)}.
\eeq

 In this form, the information from CMB maps can been extracted and $C_l^{\phi \phi}$ can be fit using $C_l^{\phi \phi} = K_{l}^{(2,2)}/{\cal K}_{l}^{(2,2)}$.   
 The signal-to-noise ratio for this estimator may be calculated as $\sqrt{\sum_l (2l+1) {\cal K}_l^{(2,2)}/24}$ where the factor of $24$ is needed to avoid over-counting of modes.  We find the total signal-to-noise for each WMAP channel is 1.61 and when the two channels are combined is 2.5.

\begin{figure}[t!]
    \begin{center}
      \includegraphics[scale=0.45]{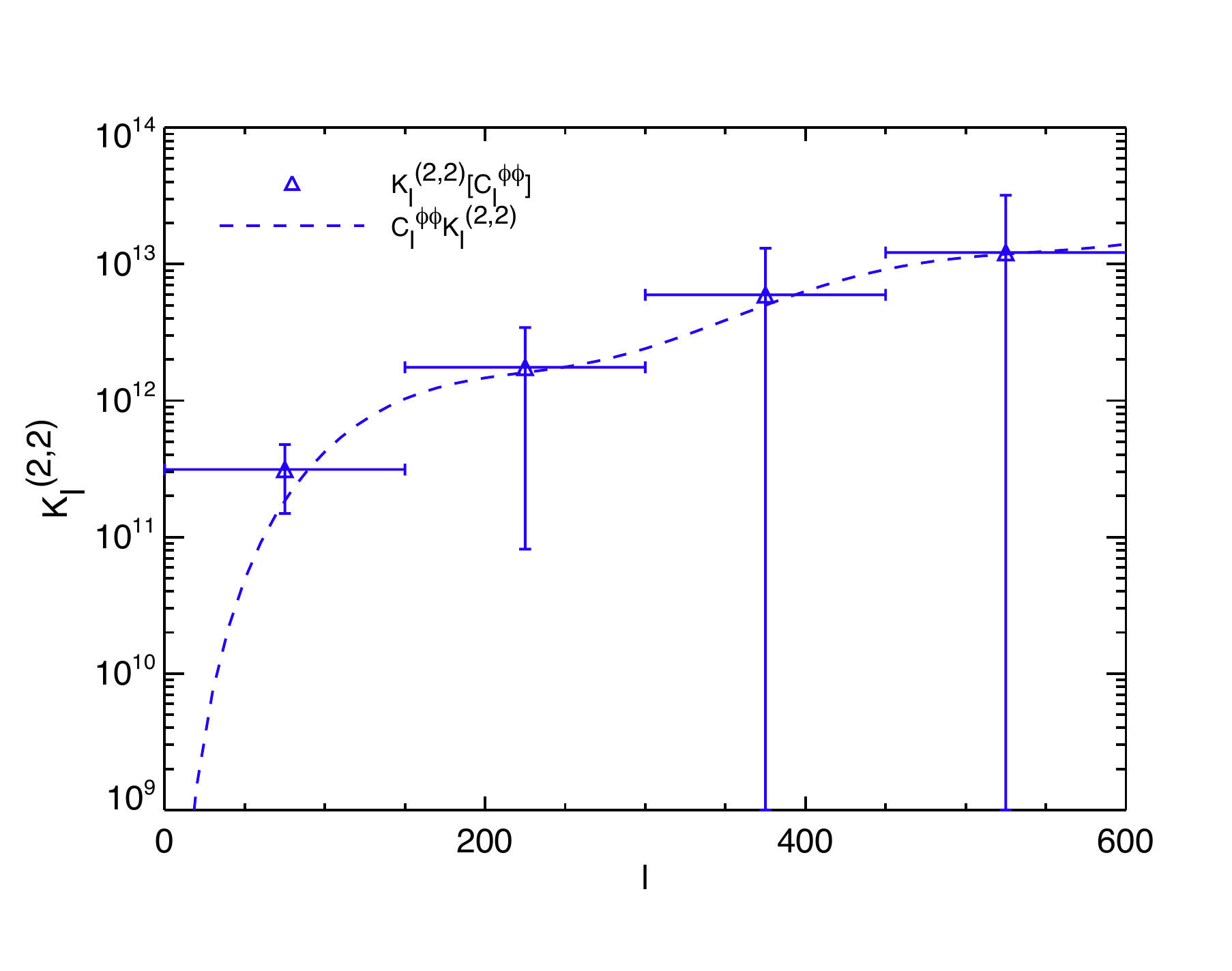} 
   \end{center}
   \caption[width=3in]{The dashed line is the analytical calculation kurtosis power spectrum $K_{l}^{(2,2)}$ for the fiducial cosmology in the WMAP W-band. The measured values are extracted from simulated lensed maps generated by {\it Lenspix} seeded by $C_l^{\phi \phi}$.}
   \label{fig:compare_sims}
\end{figure}

\section{Testing the Estimator}
\label{sec:test} 

To test this estimator we create 400 lensed and unlensed CMB temperature maps for analysis.  To create the unlensed maps, we calculate a temperature power spectrum from CAMB~(Lewis 2000) from the best-fit WMAP 7-year parameters~(Komatsu 2010).  We then use the {\it synfast} Healpix~(Gorski et al. 2005) routine to create a Gaussian map from this power spectrum and repeat until 400 maps are generated. These maps retain information from the power spectrum up to $l = 1000$ and have $n_{\rm side}=512$.   The {\it anafast} Healpix routine masks these maps with the $KQ75$ mask and produces the $a_{lm}$'s to $l = 750$.   We will refer to these $a_{lm}$'s now collectively as $a^G_{lm}$. 

In addition, we create 400 noise maps for each each of our frequency bands V and W. We generate these maps from white noise as described in~(Smidt et al. 2009) and denote the spherical harmonic coefficients as $a^N_{lm}$.  The full simulated maps combine the Gaussian maps and  noise maps in harmonic space by  $a^S_{lm} = a^G_{lm} b_l + a^N_{lm}$, where $b_l$ was obtained from the WMAP Team\footnote[1]{${\rm http://lambda.gsfc.nasa.gov/product/map/current}$} and $a^S_{lm}$ represent the spherical harmonics for the full unlensed simulation.  

Next, we generate 400 Lensed CMB maps using {\it Lenspix}~(Lewis 2008) that first creates Gaussian maps from an input temperature spectrum, then lenses the map with an input lensing power spectrum. For both spectra, we used CAMB generated fiducial temperature and lensing power spectra based on the WMAP 7-year parameters. From our maps, we obtain the lensed spherical harmonics $a^L_{lm}$ using {\it anafast}.  We then compile the full lensed simulation by adding these with the beam transfer functions and noise multipole moments as described above, utilizing the KQ75 mask.

After the Gaussian and lensed simulations are obtained, ${\cal K}_{l}^{(2,2)}$ may be extracted. Each simulation is weighted in the several combinations from which the kurtosis power spectra ${\cal K}_{l}^{(x) (2,2)}$ are formed and then added together to obtain the full estimator as shown in Eq.~\ref{eq:eexpK}.  These power spectra initially contain the Gaussian piece that must be subtracted out. Therefore, we take the mean of the kurtosis spectra obtained from the Gaussian simulations and subtract this from each kurtosis spectrum obtained from the simulated lensed maps.  To correct for the cut sky, we use the technique developed by Hivon et al. (2001) that removes masking effects using a mode-coupling matrix akin to what was done in Smidt et al. (2010).  The leftover connected pieces are averaged over yielding the theoretical ${\cal K}_l^{\phi \phi}$ for fiducial cosmology.   

The kurtosis power spectrum $K_l^{\phi \phi}$ may be computed analytically from Eq.~\ref{eq:Tperm}-\ref{eq:expK} using an unlensed temperature power spectrum and $C_l^{\phi \phi}$ obtained from CAMB.    A plot of ${\cal K}_{l}^{(2,2)}$ versus $C_l^{\phi \phi} K_{l}^{(2,2)}$ is given in Figure~\ref{fig:compare_sims}, demonstrating the estimator works well with simulated data.

\section{Analysis}
\label{sec:Analysis}
To extract ${\cal K}_l^{(2,2)}$ from data we use the raw WMAP 7-year Stokes-I sky maps for the V (60.8 GHz) and W (93.5 GHz) frequency available\footnotemark{1} from the LAMBDA website. We analyze these maps with {\it anafast} (KQ75 mask) to generate the multipole moments for each frequency band out to $l_{\rm max} = 750$.  We make no separate correction for unresolved point sources as their contamination for lensing measurements has been shown to be negligible (Smith et al. 2008).  The correction for the cut sky is handled as described above.

\begin{figure}[t]
    \begin{center}
      \includegraphics[scale=0.45]{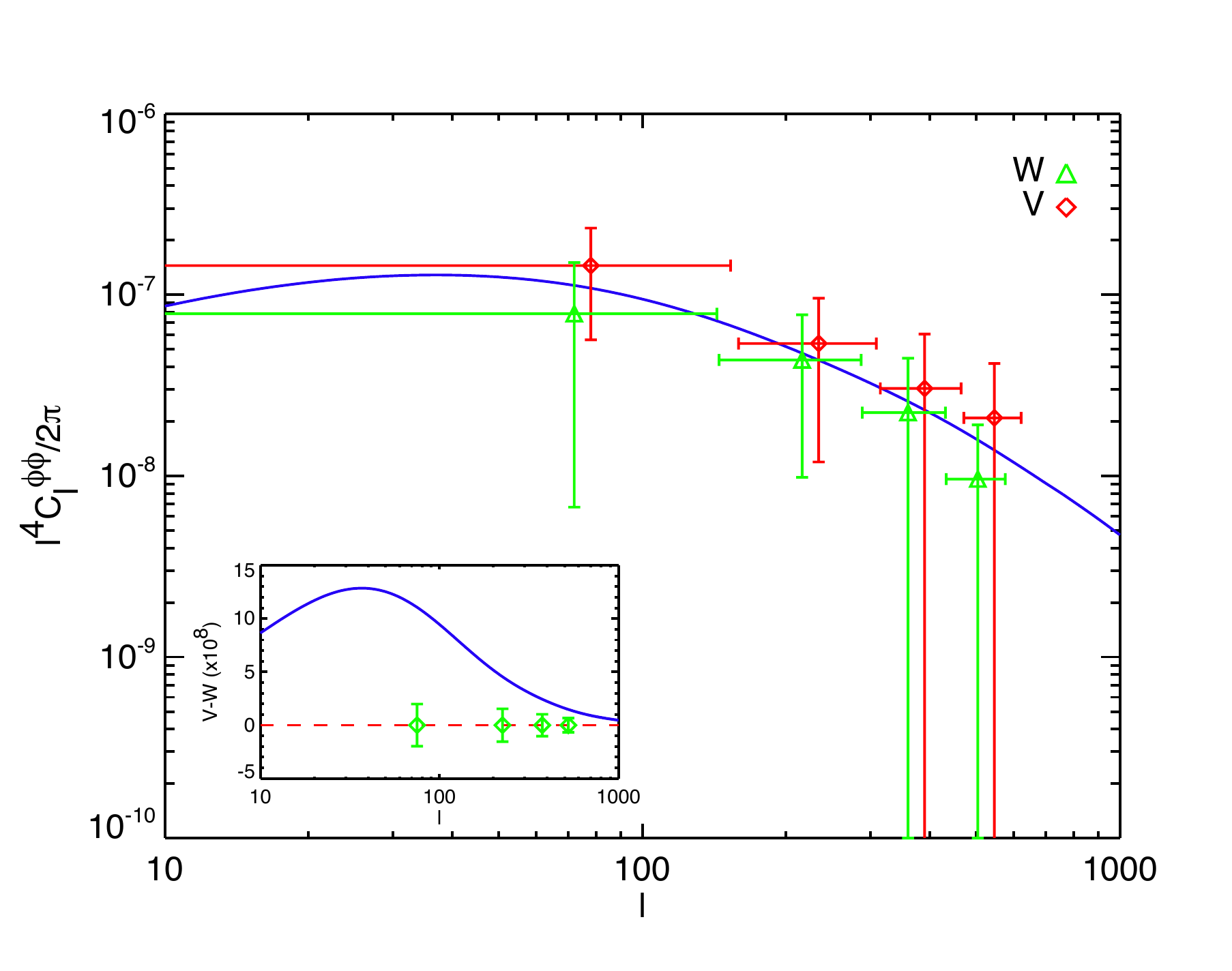} 
   \end{center}
   \caption[width=3in]{The constraints on $C_l^{\phi\phi}$ using the ${\cal K}^{(2,2)}$ estimator.  In the larger plot, the red squares are the constraints by the V-band and the green circles are for the W-Band.  The result of the null test is given in the smaller plot. These error bars represent 1$\sigma$ errors.}
   \label{fig:5fit}
\end{figure}

These data maps are then weighted and ${\cal K}_l^{(2,2)}$  is extracted with the Gaussian piece removed..  With ${\cal K}_l^{(2,2)}$, $ C_l^{\phi \phi} = {\cal K}_l^{(2,2)}/K_l^{(2,2)}$ can now be constrained.  To constrain each bin, we minimize $\chi^2$ making use of the covariance between bands and $l$ bins as described in Smidt et al. (2009).  The covariance matrix was computed and used to capture correlations between each $l$-mode. A plot of these constraints is given in Fig~\ref{fig:5fit}.  For this fit, ${\cal K}_l^{(2,2)}$ is binned in $l$ with $\delta l = 150$.

\begin{table}[t!]
\begin{center}
 \begin{tabular}{ @{} cccc  @{}}
 \hline
Params. & WMAP7 & WMAP7+$A_L$ & WMAP7+$A_L$+$C_l^{\phi \phi}$ \\
 \hline
$10^3\Omega_bh^2$ & $22.51 \pm 0.62$ & $22.59 \pm 0.63$ &
$22.60 \pm 0.58$  \\
$10^2\Omega_{DM}h^2$ & $11.08 \pm 0.57$ & $11.04 \pm 0.54$ & $11.09
\pm 0.54$  \\
$\tau$ & $0.089 \pm 0.016$ & $0.090 \pm 0.015$ & $0.089 \pm 0.015$  \\
$n_s$ & $0.967 \pm 0.015$ & $0.968 \pm 0.014$ & $0.968 \pm 0.014$  \\
$\Omega_\Lambda$ & $0.734 \pm 0.031$ & $0.737 \pm 0.028$ & $0.735
\pm 0.027$  \\
$Age/Gyr$ & $13.8 \pm 0.14$ & $13.7 \pm 0.14$ & $13.7 \pm 0.13$  \\
$H_0^{1}$ & $71.0 \pm 2.7$ & $71.3 \pm 2.5$ & $71.1\pm 2.4$  \\
\hline
${\bf A_L}$ & ${\bf 1.0}$ & ${\bf 0.87 \pm 1.05}$ & ${\bf 0.97 \pm 0.47}$  \\
 \hline
\end{tabular}
 \caption{Constraints on the cosmological parameters using CosmoMC. From left moving right: WMAP 7-year data only; WMAP7 with $A_L$ allowed to run; WMAP7 with the constraints on $C_l^{\phi \phi}$ coming from our ${\cal K}_l^{(2,2)}$ estimator from the combined W-band and V-band data.  The $^{1}$ Units on $H_0$ are (km s$^{-1}$ Mpc$^{-1}$). }
 \end{center}
  \label{tab:results}
\end{table}

To avoid a false detection, we perform a null test measuring ${\cal K}_{l}^{(2,2)}$ on the difference of temperature maps. Here, we extract our estimator from the difference of the V and W frequency band maps. The reason for calculating this difference is that the lensing signal should be the same in both maps, and therefore a subtraction of the maps should remove any lensing signal present.  The difference is performed both on the raw Stokes-I WMAP 7-year data maps, as well as the Gaussian simulations described above. Figure~\ref{fig:5fit} inset shows no lensing is detected from these differenced maps.  We also tested our estimator against unlensed maps seeded with primordial non-Gaussinainity of $f_{\rm NL} = 100$ ( Elsner \& Wandelt 2009) and found these are consistent with the Gaussian case for our estimator.   This agrees with calculations from Lesgourgues et al. 2005.

With constraints on $C_l^{\phi\phi}$, we use CosmoMC~(Lewis and Bridle 2002) to constrain the lensing amplitude $A_L$, as compared to that from the WMAP7 measurements of the temperature and E-mode polarization spectra $C_l^{\Theta \Theta}$, $C_l^{\Theta E}$, and $C_l^{EE}$ alone. The parameter $A_L$ can be thought of as a measure of the degree of lensing in the CMB, where $A_L = 1$ represents a universe with the expected amount of lensing signal and $A_L = 0$ is the case with no lensing.  We sample the following seven-dimensional set of cosmological parameters: the baryon and cold dark matter densities $\Omega_b h^2$ and $\Omega_{DM} h^2$, the ratio of the sound horizon to the angular diameter distance at the decoupling, $\theta_s$, the scalar spectral index $n_s$, the overall normalization of the spectrum $A_s$ at $k = 0.002$ Mpc$^{-1}$, and the optical depth to reionization, $\tau$.

\section{Results and Discussion}
\label{sec:final}

Fig.~\ref{fig:5fit} shows the scale dependent constraints on the lensing power spectrum obtained from the ${\cal K}_l^{(2,2)}$ estimator.  We find the lensing signal to be compatible with the fiducial expectation.  To our knowledge, this result provides the first direct constraints on $C_l^{\phi \phi}$ using CMB data encoding information for the matter distribution of the universe back to $z\sim1100$.   Furthermore,  this measurement does not appear to be biased by instrumental effects since the null test described above, is compatible with zero.

The results for measuring the lensing amplitude $A_L$ using the CosmoMC analysis described in Section~\ref{sec:Analysis} are summarized in Table~\ref{tab:results}.  We first reproduce values for the cosmological parameters consistent with Komatsu et al. (2010) for the case with no constraints on $C_l^{\phi\phi}$ and $A_L = 1$.  Next we carry out the same run, without a constraint on $C_l^{\phi\phi}$, this time allowing $A_L$ to vary and find that WMAP 7-year data alone is consistent with an unlensed universe with $A_L = 0.87 \pm 1.05$. Finally, we find $A_L = 0.96 \pm 0.60$ and  $A_L = 1.05 \pm 0.69$ when our constraints on the lensed power spectrum using the W and V frequency bands respectively are added to the WMAP 7-year data and is $A_L = 0.97 \pm 0.47$ when the W and V frequency bands are combined..

The amplitude of the $C_l^{\phi \phi}$ reported here is consistent with the
naive expectation based on a Fisher matrix estimate that suggests a
measurement no better than $2.5\sigma$. A variety of effects could be
aiding the detection; our estimator is sensitive  
to the full trispectrum while a Fisher matrix estimate based on the
reduced trispectrum with one pairing could underestimate the expected
signal-to-noise ratio (see, Fig~3 of Hu 2001b). Effects such as the
correlation between lensing and secondary anisotropies, such as the
Sunyaev-Zel'dovich effect, could enhance the signal in the lensing
trispectrum (Cooray \& Kesden 2003). While the full WMAP dataset will
slightly improve the measurement we report here, a confirmation of our
result showing a detection of the lensing power spectrum will come
from Planck data, which is expected to make a larger than 60$\sigma$
detection of $A_L$ (Hu 2001b).

As the measurement of $C_l^{\phi \phi}$ becomes more precise, it will provide tighter constraints on the cosmological parameters than can be obtained using temperature and polarization information alone.    One obvious future target is a measurement of the sum of the neutrino masses, leading to a direct cosmological determination of the neutrino mass hierarchy~(Kaplinghat et al. 2003; Lesgourgues et al. 2006).   A the lensing power spectrum further constrains the dark energy equation of state $w$ as well as early dark energy models in general (Kaplinghat et al. 2003; Joudaki in Prep.). The ultimate goal of CMB polarization measurements is the primordial gravitational wave signal in the so-called B-modes of polarization~(Baumann et al. 2009; Bock et al. 2009). The signal however is confused by the lensing effect that converts a small fraction of dominant scalar polarization in E-modes to B-modes~(Kesden et al. 2002; Knox \& Song 2002,Zaldarriaga, \& Seljak 1998).  The fast estimator we have presented here for temperature maps can be generalized for CMB polarization maps (Munshi et al. in prep) and can be used to de-lens CMB B-modes to separate the signal from primordial gravitational waves and lensing of E-modes.
\acknowledgments 
We are grateful for comments provided by Eiichiro Komatsu and Kendrick Smith.  DM acknowledges support from STFC grant ST/G002231/1 at School of Physics and
Astronomy at Cardiff University. This work was also supported by NSF CAREER AST-0645427 and NASA NNX10AD42G.

.

\end{document}